\documentclass[
aps,%
12pt,%
final,%
notitlepage,%
oneside,%
onecolumn,%
nobibnotes,%
nofootinbib,%
superscriptaddress,%
showpacs,%
centertags]%
{revtex4}
\usepackage{graphicx}
\usepackage{amssymb}
\usepackage{amsmath}
\usepackage{amsfonts}
\usepackage{caption}
\usepackage{mitpress}
\usepackage{vmargin}

\def\beq{\begin{eqnarray}}
\def\eeq{\end{eqnarray}}

\def\bg{\begin{gather}}
\def\eg{\end{gather}}
\def\m{|m|}

\def\l{\left}
\def\r{\right}

\begin{document}
\title
{Eigenfunction expansions in the imaginary Lobachevsky space
}

\author{Yu. A. Kurochkin}
\email {yukuroch@bas-net.by}
\author{V.S. Otchik}
\email{votchyk@tut.by}
\affiliation{B.I. Stepanov Institute of Physics, Minsk, Belarus}
\author{D.R. Petrosyan}
\email{petrosyan@theor.jinr.ru}
\affiliation{Joint Institute for Nuclear Research, Dubna, Russia}
\author{G.S. Pogosyan}
\email{pogosyan@theor.jinr.ru}
\affiliation{Joint Institute for Nuclear Research, Dubna, Russia}
\affiliation{Departamento de Matematicas, CUCEI,
Universidad de Guadalajara,  Guadalajara, Mexico}
\affiliation{Yerevan State University, Yerevan, Armenia}

\begin{abstract}
Eigenfunctions of the Schr\"odinger equation with the Coulomb potential in the imaginary Lobachevsky space are studied in two coordinate systems admitting solutions in terms of hypergeometric functions. Normalization and coefficients of mutual expansions for some sets of solutions are found.\end{abstract}

 \maketitle

Geometry of spaces of constant curvature with (pseudo)-orthogonal group of motion provides a natural framework for treatment of numerous physical problems. For example, the velocity space in relativity is the Lobachevsky space, and the imaginary Lobachevsky space corresponds to an unphysical region of momentum variables, which is also of interest for the study of scattering processes. Metrics of spaces of constant curvature appear as simplest solutions of Einstein equations and are used as a background in models aimed at the study of various effects of curvature in quantum theory. Derivation of eigenfunction expansions is an essential element of such study.

Eigenfunction expansions associated with invariant operators on various hyperboloids in arbitrary dimensions were derived in \cite{LNR} and in special case of four-dimensional pseudo-Euclidean space in \cite{Zm}. Authors of these works used the canonical subgroup reduction and separation of variables associated with it.
Harmonic analysis in homogeneous spaces based on the methods of integral geometry was developed in \cite{GGV}. These methods were applied in \cite{VS} to obtain eigenfunction expansions in several coordinate systems in the Lobachevsky space (the upper sheet of a double-sheeted hyperboloid), and also in \cite{KaM}, where
separable bases on one and two sheeted hyperboloids were studied. (In \cite{GGV, KaM} the imaginary Lobachevsky space is defined as the single sheeted hyperboloid with geometrically opposed points identified).

Study of quantum mechanical problems in spaces of constant curvature was initiated by Schr\"odinger \cite{Schr}, who solved the Kepler--Coulomb problem on a three-dimensional sphere. In the Lobachevsky space, this problem was solved by Infeld and Schild \cite{Inf}. Since then, a considerable number of works
concerning the Kepler--Coulomb problem in these two spaces (see e.g. \cite{Hig}--\cite{PETPOG4} and references therein) has appeared.

On the other hand, C. Grosche \cite{Gro} in his development of path integral in the imaginary Lobachevsky space incorporated in it an analog of the Coulomb potential and presented solution of thus formulated Kepler--Coulomb problem in this space. For all three spaces of constant curvature, symmetry of this problem is essentially common \cite{BKOa}. However, much less is known about eigenfunctions of the problem in the imaginary Lobachevsky space than the other two. In this paper we consider the Kepler--Coulomb problem in two coordinate systems of imaginary Lobachevsky space admitting separation of variables and solutions in terms of hypergeometric functions.

In terms of ambient pseudo-Euclidean space, the imaginary Lobachevsky space
is the single-sheet hyperboloid $(\mu = 0,1,2,3)$
$$
x_{\mu} x_{\mu} = {\bf x}^{2} - x_0^{2} = R^{2},
\qquad
{\bf x} = \sqrt{x_1^2 + x_2^2 +x_3^2}.
$$

The Schr\"odinger equation for the Kepler problem in this space is written as
\begin{equation}\label{Scheq}
H\Psi=E\Psi, \quad H=-\frac12\Delta+V,
\end{equation}
where $\Delta$ is the Laplace operator on the hyperboloid, and the Coulomb
potential $V$ is \cite{Gro}:
$$
V=-\frac{\alpha}{R}\frac{x_0}{|\bf x|},  \qquad (\alpha>0).
$$

The spherical coordinate system in the imaginary Lobachevsky space can be introduced by the relations
$$
x_{0} = R \sinh\tau,\,
x_{1}=R \cosh\tau\sin\theta \cos\phi, \,
x_{2} = R \cosh\tau \sin\theta \sin\phi,\,
x_{3} = R \cosh\tau \cos\theta,
$$
\begin{equation}
\label{imls}
 -\infty<\tau<\infty, \quad 0\leq\theta\leq\pi, \quad 0\leq\phi<2\pi.
\end{equation}
In the coordinates $\tau,\, \theta,\, \phi$, the line element is
$$
\frac{ds^2}{R^2} = [d\tau^2 - \cosh^2\tau (d\theta^2 + \sin^2\theta d\phi^2)],
$$
and the Hamiltonian (\ref{Scheq}) is
\begin{equation}\label{Hms}
H=-\frac{1}{2R^2}\left[\frac{\partial^2}{\partial \tau^2}+2\tanh\tau\frac{\partial}{\partial\tau}-\frac{1}{\cosh^2\tau} \left(\frac{\partial^2}{\partial\theta^2}+\cot\theta\frac{\partial}{\partial\theta}
+\frac{1}{\sin^2\theta}\frac{\partial^2}{\partial\phi^2}\right)\right] -\frac{\alpha}{R}\tanh\tau.
\end{equation}
Writing the wave function in the form
\begin{equation}\label{sv}
\Psi=S(\tau)Y_{\ell m}(\theta,\phi),
\end{equation}
where $Y_{\ell m} (\theta,\phi)$ are spherical harmonics \cite{Var}, we obtain an equation for the radial
function $S(\tau)$
\begin{equation}\label{req}
\frac{d^2S}{d\tau^2}+2\tanh\tau\frac{dS}{d\tau} +
\left[\frac{\ell(\ell+1)}{\cosh^2\tau}+2\alpha
R\tanh\tau+2ER^2\right]S=0.
\end{equation}
Denoting $z=(1+\tanh\tau)/2=(1+e^{-2\tau})^{-1},$ we get from (\ref{req}) equation
\begin{equation}\label{requ}
\frac{d^2S}{dz^2}+\frac{1}{z(z-1)}\left[-\ell(\ell+1)+\frac{ER^2+\alpha
R}{2(z-1)}-\frac{ER^2-\alpha R}{2z}\right]S=0,
\end{equation}
which can be solved in terms of hypergeometric functions. Setting further
\begin{equation}\label{f}
S(\tau)=(\cosh\tau)^{-1}f(\tau), \vspace{-0.1cm}
\end{equation}
we arrive at equation \vspace{-0.1cm}
\begin{equation}\label{reqf}
\frac{d^2f}{d\tau^2}+ \left[\frac{\ell(\ell+1)}{\cosh^2\tau}+2\alpha
R(\tanh\tau-1)+\Lambda\right]f=0,
\end{equation}
where $\Lambda=2ER^2+2\alpha R-1$. Equation (\ref{reqf}) has the form of the one-dimensional Schr\"odinger equation with the
Rosen--Morse potential \cite{RM}. For the discrete energy states, normalization of wave functions for this potential was accomplised
in the work \cite{Nie}. Here we use the method of Titchmarsh \cite{Tit}, which allows one to obtain normalization constants for
both discrete and continuous states. From now on we consider $\Lambda$ as a complex variable. Let $f_1(\tau)$ and $f_2(\tau)$
be two linearly independent solutions of equation (\ref{reqf}) such that for
$\mathrm{Im} \Lambda > 0$
\begin{equation}\label{pro}
f_1(\tau)\in L^2(-\infty,\,0), \quad f_2(\tau)\in
L^2(0,\,+\infty), \vspace{-0.1cm}
\end{equation}
where $L^2(a,\,b)$ denotes the space of square integrable functions on
the interval $(a,\,b)$. Consider now a function \vspace{-0.1cm}
\begin{equation}\label{ph}
\Phi(\tau,\Lambda)=\frac{f_2(\tau)}{W(f_1,f_2)}
\int_{-\infty}^{\tau}f_1(\tau^{\prime})\varphi(\tau^{\prime})d\tau^{\prime}
+\frac{f_1(\tau)}{W(f_1,f_2)}
\int_{\tau}^{\infty}f_2(\tau^{\prime})\varphi(\tau^{\prime})d\tau^{\prime}.
\end{equation}
Here $\varphi(\tau)$ is an arbitrary function square integrable on
$(-\infty,\,+\infty)$ and $W(f_1,f_2)$ is the Wronskian \vspace{-0.1cm}
$$
W(f_1,f_2)=f_1\frac{df_2}{d\tau}-f_2\frac{df_1}{d\tau}.
$$
Then the eigenfunction expansion of $\varphi(\tau)$ associated with equation (\ref{reqf}) is
\begin{equation}\label{gexp}
\varphi(\tau)=\frac{1}{\pi}\int_{-\infty}^{\infty} [-\mathrm
{Im}\Phi(\tau,\Lambda)]d\Lambda + \sum_n \mathrm {Res}\,\Phi(\tau,\Lambda_n), \vspace{-0.1cm}
\end{equation}
where $\mathrm{Res}\,\Phi(\tau,\Lambda_n)$ are residues of $\Phi(\tau,\Lambda)$ in the complex $\Lambda$ plane.

First we consider the case of free motion $\alpha=0$. In this case, the equation (\ref{reqf}) reduces to
\begin{equation}\label{reqff}
\frac{d^2f}{d\tau^2}+ \left[\frac{\ell(\ell+1)}{\cosh^2\tau}+\Lambda\right]f=0,
\end{equation}
with $\Lambda=2ER^2-1$.
Consider two solutions of (\ref{reqff}) satisfying conditions (\ref{pro})
\begin{equation}\label{f1f}
f_1(\tau)=(2\cosh\tau)^{ip}{}_2F_1\left(\ell+1-ip,
-\ell-ip;1-ip;\,(1+e^{-2\tau})^{-1}\right),
\end{equation}
\begin{equation}\label{f2f}
f_2(\tau)=(2\cosh\tau)^{ip}{}_2F_1\left(\ell+1-ip,
-\ell-ip;1-ip;\,(1+e^{2\tau})^{-1}\right),
\end{equation}
where $p=\sqrt{\Lambda}$. The Wronskian of solutions (\ref{f1f}) and (\ref{f2f}) is
\begin{equation}\label{wf}
W(f_1,f_2)=-\frac{2\left[\Gamma\left(1-ip\right)\right]^2}
{\Gamma\left(\ell+1-ip\right)\Gamma\left(-\ell-ip\right)}.
\end{equation}
By using properties of hypergeometric functions \cite{Erd} one can easily check that
\begin{eqnarray}
f_1^*(\tau)=f_1(\tau), \quad f_2^*(\tau)=f_2(\tau) \quad \,\,
\mbox{for} \, \Lambda<0; \label{fcm}\\
\begin{array}{l}
{\displaystyle f_1^*(\tau)=\frac{\Gamma\left(\ell+1-ip\right)
\Gamma\left(-\ell-ip\right)}{\Gamma\left(1-ip\right)
\Gamma\left(-ip\right)}f_2(\tau), }  \\[4mm]
 {\displaystyle f_2^*(\tau)=\frac{\Gamma\left(\ell+1-ip\right)
\Gamma\left(-\ell-ip\right)}{\Gamma\left(1-ip\right)
\Gamma\left(-ip\right)}f_1(\tau)}
\end{array}
  \quad \mbox{for} \,
\Lambda>0. \label{fcp}
\end{eqnarray}
Using equations (\ref{wf}) -- (\ref{fcp}) we find the discontinuity of the function
$\Phi(\tau,\Lambda)$ (\ref{ph}) on the real axis of complex $\Lambda$ plane:
\begin{equation}
\mathrm {Im}\Phi(\tau,\Lambda)=0 \quad \mbox{for} \, \Lambda<0;
\end{equation}
\vspace{1mm}
\begin{equation}
\mathrm {Im}\Phi(\tau,\Lambda)=-\frac{1}{4p}\left[
f_1^*(\tau)
\int_{-\infty}^{\infty}f_1(\tau^{\prime})\varphi(\tau^{\prime})
d\tau^{\prime} + f_2^*(\tau) \int_{-\infty}^{\infty}
f_2(\tau^{\prime}) \varphi(\tau^{\prime})d\tau^{\prime}\right]
\quad \mbox{for} \, \Lambda>0.
\end{equation}

The function $\Phi(\tau,\Lambda)$ has poles at the points $-i\sqrt{\Lambda}=n$, where $n=1,2,\ldots, \ell$.
Since $f_1(\tau)=(-1)^{\ell-n}f_2(\tau)$ for these values of $\Lambda$, the residues of $\Phi(\tau,\Lambda)$
are found to be
\begin{equation}
\mathrm {Res}\,\Phi(\tau,\Lambda_n)=\frac{(\ell+n)!\,n}{(\ell-n)!(n!)^2}f_{1n}(\tau) \int_{-\infty}^{\infty}
f_{1n}(\tau^{\prime})\varphi(\tau^{\prime})d\tau^{\prime},
\end{equation}
where
\begin{equation}
f_{1n}(\tau)=(2\cosh\tau)^{-n}{}_2F_1\left(n+\ell+1,
n-\ell;n+1;\,(1+e^{-2\tau})^{-1}\right).
\end{equation}

Thus eigenfunction expansion (\ref{gexp}) of an arbitrary function $\varphi$ in case of equation (\ref{reqff})
takes the form
\begin{equation}
\label{expf}
\begin{array}{c}
\varphi(\tau)={\displaystyle\frac{1}{4\pi}\int_0^{\infty} \frac{d\Lambda}{\sqrt{\Lambda}}\left[f_1^*(\tau)
\int_{-\infty}^{\infty}f_1(\tau^{\prime}) \varphi(\tau^{\prime})
d\tau^{\prime} + f_2^*(\tau)
\int_{-\infty}^{\infty} f_2(\tau^{\prime})
\varphi(\tau^{\prime})d\tau^{\prime}\right] }
\\[5mm]
+ {\displaystyle \sum_{n=1}^\ell \frac{(\ell+n)!n} {(\ell-n)!(n!)^2}f_{1n}(\tau)
\int_{-\infty}^{\infty}f_{1n}(\tau^{\prime})\varphi(\tau^{\prime})d\tau^{\prime}. }
\end{array}
\end{equation}
Expressions for normalized eigenfunctions of equation (\ref{reqff}) can be easily deduced from the
expansion (\ref{expf}).

Now we return to the equation (\ref{reqf}) with $\alpha>0$. Introducing notations
\begin{equation}
p=\sqrt{\Lambda}, \quad q=\sqrt{\Lambda-4\alpha R}, \quad \nu=-\frac{i}{2}(p+q),
\end{equation}
we can write solutions of (\ref{reqf}) satisfying conditions (\ref{pro}) as follows:
\begin{equation}
\label{f1c}
f_1(\tau)= (1+e^{-2\tau})^{iq/2} (1+e^{2\tau})^{ip/2} {}_2F_1\left(\nu+\ell+1, \nu-\ell; 1-iq;\,(1+e^{-2\tau})^{-1}\right),
\end{equation}
\begin{equation}
\label{f2c}
f_2(\tau)= (1+e^{-2\tau})^{iq/2} (1+e^{2\tau})^{ip/2} {}_2F_1\left(\nu+\ell+1, \nu-\ell; 1-ip;\,(1+e^{2\tau})^{-1}\right).
\end{equation}
Wronskian of the solutions (\ref{f1c}) and (\ref{f2c}) is \vspace{-0.1cm}
\begin{equation}
\label{wc}
W(f_1,f_2)=-\frac{2\Gamma\left(1-ip\right)
\Gamma\left(1-iq\right)}
{\Gamma\left(\nu+\ell+1\right)\Gamma\left(\nu-\ell\right)}.
\end{equation}
Denoting $\sqrt{4\alpha R-\Lambda}=r$ for $\Lambda<4\alpha R$ and using relations between solutions
of hypergeometric equation we find
\begin{equation}\label{fccm}
f_1^*(\tau)=f_1(\tau), \quad f_2^*(\tau)=f_2(\tau) \quad
\mbox{for} \, \Lambda<0;
\end{equation}
\vspace{2mm}
\begin{equation}\label{fcci}
\begin{array}{r}
{\displaystyle{f_1^*(\tau)=f_1(\tau)=\frac{\Gamma\left(ip\right)
\Gamma\left(1+r\right)f_2(\tau)}
{\Gamma\left(\ell+1+\frac{ip}{2}+ \frac{r}{2}\right)\Gamma\left(-\ell+\frac{ip}{2}+
\frac{r}{2}\right)}} }
\\[6mm]
{\displaystyle +\frac{\Gamma\left(-ip\right)
\Gamma\left(1+r\right)f_2^*(\tau)}
{\Gamma\left(\ell+1-\frac{ip}{2}+ \frac{r}{2}\right)\Gamma\left(-\ell-\frac{ip}{2}+
\frac{r}{2}\right)} }
\end{array}
  \quad \mbox{for} \,
0<\Lambda<4\alpha R;
\end{equation}
\vspace{2mm}
\begin{equation}\label{fccp}
\begin{array}{c}
{\displaystyle f_1(\tau)=\frac{\Gamma\left(ip\right)
\Gamma\left(1-iq\right) f_2(\tau)}
{\Gamma\left(\ell+1+\frac{ip}{2}-
\frac{iq}{2}\right)\Gamma\left(-\ell+\frac{ip}{2}-
\frac{iq}{2}\right)}
+\frac{\Gamma\left(-ip\right)
\Gamma\left(1-iq\right) f_2^*(\tau)}
{\Gamma\left(\ell+1-\frac{ip}{2}-
\frac{iq}{2}\right)\Gamma\left(-\ell-\frac{ip}{2}-
\frac{iq}{2}\right)}, }
\\[6mm]
{\displaystyle f_2(\tau)=\frac{\Gamma\left(1-ip\right)
\Gamma\left(iq\right) f_1(\tau)}
{\Gamma\left(\ell+1-\frac{ip}{2}+
\frac{iq}{2}\right)\Gamma\left(-\ell-\frac{ip}{2}+
\frac{iq}{2}\right)}
+\frac{\Gamma\left(1-ip\right)
\Gamma\left(-iq\right) f_1^*(\tau)}
{\Gamma\left(\ell+1-\frac{ip}{2}-
\frac{iq}{2}\right)\Gamma\left(-\ell-\frac{ip}{2}-
\frac{iq}{2}\right)} } \vspace{0.1cm}
\\[3mm] \mbox{for} \, \Lambda>4\alpha R.
\end{array}
\end{equation}
By the use of equations (\ref{wc}) -- (\ref{fccp}) we obtain:
\begin{equation}\label{imm}
\mathrm{Im}\Phi(\tau,\Lambda)=0 \quad \mbox{for} \, \Lambda<0;
\end{equation}
\vspace{2mm}
\begin{equation}\label{imi}
\begin{array}{c}
\mathrm
{Im}\Phi(\tau,\Lambda)={\displaystyle -\frac{\sinh\pi p\,\left|\ell+1-
\frac{ip}{2}+ \frac{r}{2}\right|^2\left|-\ell-\frac{ip}{2}+
\frac{r}{2}\right|^2
}{4\pi\left[\Gamma\left(1+\frac{r}{2}\right)\right]^2}} {\displaystyle f_1(\tau)
\int_{-\infty}^{\infty}f_1(\tau^{\prime}) \varphi(\tau^{\prime})
d\tau^{\prime} }\vspace{0.2cm} \\ \mbox{for} \,\, 0<\Lambda<4\alpha R; \quad  \quad \quad \quad
\end{array}
\end{equation}
\vspace{2mm}
\begin{equation}\label{imp}
\begin{array}{c}
\mathrm
{Im}\Phi(\tau,\Lambda)\!=\! {\displaystyle -\frac{\sinh\pi p\,\sinh\pi q} {4\sinh^2\frac{\pi}{2}\left(p+q\right)} }
{\displaystyle \left[\frac{1}{q} f_1^*(\tau) \!
\int_{-\infty}^{\infty} \! f_1(\tau^{\prime}) \varphi(\tau^{\prime})
d\tau^{\prime} \! + \! \frac{1}{p}f_2^*(\tau) \!
\int_{-\infty}^{\infty}\!  f_2(\tau^{\prime})
\varphi(\tau^{\prime})d\tau^{\prime}\right] } \vspace{0.2cm}\\
\mbox{for} \,
\Lambda>4\alpha R. \quad \quad \quad
\end{array}
\end{equation}

The function $\Phi(\tau,\Lambda)$ (\ref{ph}) has poles at the points \vspace{-0.1cm}
\begin{equation}\label{nun}
\nu=-i\left(\sqrt{\Lambda}+\sqrt{\Lambda-4\alpha R}\right)/2=n,
\end{equation}
where $n=\delta,\,\delta+1,\ldots, \ell$, and $\delta=\max\{[\sqrt{\alpha R}],\,1\}$.
For these values of $\nu$ we have
$$
f_2(\tau)=(-1)^{n-l}\frac{\Gamma(n-\sigma+1)\Gamma(\ell+\sigma+1)}
{\Gamma(n+\sigma+1)\Gamma(\ell-\sigma+1)}f_1(\tau), \quad \sigma=\frac{\alpha R}{n}.
$$
Thus we find
\begin{equation}\label{resc}
\mathrm {Res}\Phi(\tau,\Lambda_n)=\frac{(\ell+n)!\Gamma(\ell+\sigma+1)(n^2-\sigma^2)}
{(\ell-n)!\Gamma(\ell-\sigma+1)[\Gamma(n+\sigma+1)]^2n}f_{1n}(\tau) \int_{-\infty}^{\infty}f_{1n}(\tau^{\prime})\varphi(\tau^{\prime})d\tau^{\prime},
\end{equation}
where
\begin{equation}
f_{1n}(\tau)=e^{\sigma\tau}(2\cosh\tau)^{-n}{}_2F_1\left(n+\ell+1,
n-\ell; n+\sigma+1; \,(1+e^{-2\tau})^{-1}\right).
\end{equation}

From equation (\ref{nun}) follows expression for discrete energy levels
\begin{equation}\label{en}
E_n=-\frac{\alpha^2}{2n^2}-\frac{n^2-1}{2R^2}.
\end{equation}

Now we can write the eigenfunction expansion associated with the equation (\ref{reqf}) for an arbitrary
function $\varphi(\tau)$
\begin{eqnarray}
\varphi(\tau)
&=&
\int_0^{4\alpha R}d\Lambda \frac{\sinh\pi p\,\left|\ell+1-
\frac{ip}{2}+ \frac{r}{2}\right|^2\left|-\ell-\frac{ip}{2}+
\frac{r}{2}\right|^2
}{4\pi^2\left[\Gamma\left(1+\frac{r}{2}\right)\right]^2}f_1(\tau)
\int_{-\infty}^{\infty}f_1(\tau^{\prime}) \varphi(\tau^{\prime})
d\tau^{\prime}
\nonumber
\\[3mm] \label{exp}
&+& \frac{1}{4\pi}\int_{4\alpha R}^{\infty} d\Lambda \frac{\sinh\pi p\,\sinh\pi q} {\sinh^2 \frac{\pi}{2}(p+q)}
\Biggl[\frac{1}{q} f_1^*(\tau)
\int_{-\infty}^{\infty}f_1(\tau^{\prime}) \varphi(\tau^{\prime})
d\tau^{\prime}
\\[3mm]
&+& \frac{1}{p}f_2^*(\tau)
\int_{-\infty}^{\infty} f_2(\tau^{\prime})
\varphi(\tau^{\prime})d\tau^{\prime}\Biggr] + \,
 \sum_{n=\delta}^\ell \frac{(\ell+n)!\Gamma(\ell+\sigma+1)(n^2-\sigma^2)} {(\ell-n)!\Gamma(\ell-\sigma+1)[\Gamma(n+\sigma+1)]^2n}f_{1n}(\tau)
\nonumber
\\[3mm]
&\times&
\int_{-\infty}^{\infty}f_{1n}(\tau^{\prime})\varphi(\tau^{\prime})d\tau^{\prime}. \nonumber
\end{eqnarray}
Note that only $f_1(\tau)$ enters the expansion for the interval $0<\Lambda<4\alpha R$ of the continuous spectrum.
We can see from (\ref{f1c}) and (\ref{exp}) that the normalized eigenfunction for this interval can be written as
\begin{equation}\label{nf0c}
{}^0\!f_{\Lambda}(\tau)={}^0N_{\Lambda}(1+e^{-2\tau})^{-r/2} (1+e^{2\tau})^{ip/2}
{}_2F_1\left(\nu+\ell+1, \nu-\ell; 1+r;\,(1+e^{-2\tau})^{-1}\right),
\end{equation}
where $\nu=(r-ip)/2$, and
$$
{}^0N_{\Lambda}=\frac{\sqrt{\sinh\pi p}\left|\ell+1-
\frac{ip}{2}+ \frac{r}{2}\right|\left|-\ell-\frac{ip}{2}+
\frac{r}{2}\right|
}{2\pi\Gamma\left(1+\frac{r}{2}\right)}.
$$
For $\Lambda>4\alpha R$, two eigenfunctions enter the expansion (\ref{exp}). The normalized functions are
\begin{equation}\label{nf1c}
{}^1\!f_{\Lambda}(\tau)= {}^1N_{\Lambda}(1+e^{-2\tau})^{iq/2} (1+e^{2\tau})^{ip/2}
{}_2F_1\left(\nu+\ell+1, \nu-\ell; 1-iq;\,(1+e^{-2\tau})^{-1}\right),
\end{equation}
\begin{equation}\label{nf2c}
{}^2\!f_{\Lambda}(\tau)= {}^2N_{\Lambda}(1+e^{-2\tau})^{iq/2} (1+e^{2\tau})^{ip/2}
{}_2F_1\left(\nu+\ell+1, \nu-\ell; 1-ip;\,(1+e^{2\tau})^{-1}\right),
\end{equation}
where
$$
{}^1N_{\Lambda}=\frac{\sqrt{\sinh\pi p\,\sinh\pi q}}{2\sqrt{\pi q}\,\sinh\frac{\pi}{2}(p+q)},
\quad
{}^2N_{\Lambda}=\frac{\sqrt{\sinh\pi p\,\sinh\pi q}}{2\sqrt{\pi p}\,\sinh\frac{\pi}{2}(p+q)}.
$$

Normalized eigenfunctions for discrete states are
\begin{equation}\label{ndfc}
f_{nl}(\tau)=N_{n\ell}e^{\sigma\tau}(2\cosh\tau)^{-n}{}_2F_1\left(n+\ell+1,
n-\ell;n+\sigma+1;\,(1+e^{-2\tau})^{-1}\right),
\end{equation}
where
$$
N_{n\ell}=\frac{1}{\Gamma(n+\sigma+1)}\sqrt{\frac{(n^2-\sigma^2)(n+\ell)!\Gamma(\ell+\sigma+1)}
{n(\ell-n)!\Gamma(\ell-\sigma+1)}}.
$$

Functions (\ref{nf0c}) -- (\ref{ndfc}) satisfy orthogonality relations
\begin{equation}\label{or0}
\int_{-\infty}^{\infty}{}^0\!f_{\Lambda}(\tau)\,{}^0\!f_{\Lambda^{\prime}}(\tau)d\tau =
\delta(\Lambda-\Lambda^{\prime}), \quad 0<\Lambda<4\alpha R;
\end{equation}
\begin{equation}\label{or12}
\int_{-\infty}^{\infty}{}^i\!f^*_{\Lambda}(\tau)\,{}^j\!f_{\Lambda^{\prime}}(\tau)d\tau
= \delta(\Lambda-\Lambda^{\prime})\delta_{ij}, \quad i,j=1,2;\,\Lambda>4\alpha R;
\end{equation}
\begin{equation}\label{ord}
\int_{-\infty}^{\infty}f_{n\ell}(\tau)f_{n^{\prime}\ell}(\tau)d\tau = \delta_{nn^{\prime}}.
\end{equation}

Taking into account expressions (\ref{sv}) and (\ref{f}) for wave functions $\Psi(\tau,\theta,\phi)$,
we can now easily obtain solutions of the Schr\"odinger equation (\ref{Scheq}) normalized with respect
to the scalar product
\begin{equation}
(\Psi_1,\,\Psi_2)=\int\!\!\int\limits_{V}\!\!\int\Psi_1^*
\Psi_2 dV =  \int\!\!\int\limits_{V}\!\!\int\Psi_1^*
\Psi_2  R^3\cosh^2\tau\sin\theta d\tau d\theta d\phi,
\end{equation}
where integration is taken over the whole hyperboloid. In particular, the wave functions of discrete states
$\Psi_{n\ell m} (\tau,\, \theta,\, \phi)$ normalized by the condition $(\Psi_{n\ell m},\,\Psi_{n\ell m})=1$, have the form
\begin{equation}\label{psnl}
\Psi_{n\ell m} (\tau,\, \theta,\, \phi)= N_{n\ell}R^{-3/2}\, (\cosh\tau)^{-1}\! f_{n\ell} (\tau) Y_{\ell m}(\theta, \phi).
\end{equation}

In the imaginary as well as in real Lobachevsky space there exists, besides spherical coordinates,
one more separable coordinate system\footnote{As in the real Lobachevsky the Schroedinger equation with the Coulomb potential
in the imaginary Lobachevsky space admits separation of variables in four systems of coordinates.},
in which solutions of Schr\"odinger equation with the Coulomb potential can be expressed through hypergeometric functions.
In the case of Lobachevsky space, such solutions were studied in \cite{BOR}. The corresponding coordinate system from
Olevski\u \i's list \cite{Ol} was named in \cite{GPSs} elliptic-parabolic II. Solutions of the Kepler--Coulomb problem
in a similar system in the imaginary Lobachevsky space were considered in \cite{ints}.
(Eigenfunctions of the Laplace operator in the imaginary Lobachevsky space in a variety of coordinate systems were found in \cite{KaM}).
Now we proceed to find normalization coefficients for some sets of eigenfunctions. We define an analog of elliptic-parabolic II coordinate
system with the help of relations
\begin{equation}\label{t12}
t_1=\frac{|{\bf x}|-x_3}{|{\bf x}|+x_0},\;t_2=\frac{|{\bf x}|+x_3}{|{\bf x}|-x_0}.
\end{equation}
Then for $x_0+x_3 > 0$ we have $0 \leq t_1 < 1,\, t_2 > 1 $, and the coordinates of the pseudoeuclidean space are expressed as
\begin{align}\label{xt}
x_{0} &=  \frac{R(t_1+t_2-2)}{2\sqrt{(1-t_1)(t_2-1)}},\;
&x_{1} &=R \sqrt{t_1 t_2}\cos\phi, \;\nonumber\\
x_{2} &= R \sqrt{t_1 t_2} \sin\phi,\;
&x_{3} &= \frac{R(t_1+t_2-2t_1 t_2)}{2\sqrt{(1-t_1)(t_2-1)}}.
\end{align}
It should be noted that coordinates $t_1,\,t_2$ only cover one half of the one-sheeted hyperboloid.
To cover the second half: $x_0+x_3 < 0$, one can e.g. change signs of $x_0$ and $x_3$ in (\ref{xt}).
Correspondingly, we get that $t_1 > 1, \, 0 \leq t_2 < 1$.

In the coordinates $t_1,t_2,\phi$, Hamiltonian (\ref{Scheq}) takes the form
\begin{equation}
H=\frac{2}{R^2}\left[\frac{1-t_1}{t_2-t_1}\frac{\partial}
{\partial t_1}t_1(1-t_1)\frac{\partial} {\partial t_1} \! +\!
\frac{1-t_2}{t_1-t_2}\frac{\partial} {\partial
t_2}t_2(1-t_2)\frac{\partial}{\partial t_2}
\!+\!\frac{1}{4t_1t_2}\frac{\partial^2} {\partial\phi^2}\right]
\!-\!\frac{\alpha}{R}\frac{t_1+t_2-2}{t_2-t_1}.
\nonumber
\end{equation}
Substitution $\Psi=S_1(t_1)S_2(t_2)e^{im\phi}$ separates the
variables in the Schr\"odinger equation and we obtain the equations for $S_1$ and $S_2$
\begin{eqnarray}
\frac{d}{dt_1}(t_1-1)t_1\frac{dS_1}{dt_1} +
\left[\frac{ER^2-\alpha R}{2}-
\frac{m^2}{4t_1}+\frac{A}{4(t_1-1)}\right]S_1=0, \label{es1}
\\[3mm] \frac{d}{dt_2}(t_2-1)t_2\frac{dS_2}{dt_2} +
\left[\frac{ER^2+\alpha R}{2}-
\frac{m^2}{4t_2}+\frac{A}{4(t_2-1)}\right]S_2=0. \label{es2}
\end{eqnarray}

Solutions of these equations can be expressed in terms of hypergeometric functions.
Spectra of energy $E$ and of the separation constant $A$ should be found by imposing appropriate boundary conditions on these solutions.
But this task is complicated by the fact that both $E$ and $A$ enter each of the equations. In the case of free motion, when $\alpha=0$,
one finds that solutions of the equations (\ref{es1}) and (\ref{es2}) cannot be simultaneously finite for any values of $E$ and $A$.
On the other hand, the existence of the discrete energy spectrum (see (\ref{en})) was demonstrated by solving the problem
in the spherical coordinates. This means that for the discrete values of energy, the separation constant $A$ may have continuous spectrum.
Now we will find corresponding wave functions and their normalization coefficients.

Let us denote a solution of any of the equations (\ref{es1}) and (\ref{es2}) by $S_{EA}(t)$ with
given values $E$ and $A$. Then we have
\begin{equation}\label{gi}
\begin{array}{c}
{\displaystyle\int_a^b\left[2R^2(E-E^{\prime})+(A-A^{\prime})(t-1)^{-1}\right] S_{EA}S_{E^{\prime}A^{\prime}}dt } \\[3mm]
=  -{\displaystyle \left. 4\left[t(t-1)\left(S_{E^{\prime}A^{\prime}}\frac{dS_{EA}}{dt}- S_{EA}
\frac{dS_{E^{\prime}A^{\prime}}}{dt}\right)\right]\right|_a^b.}
\end{array}
\end{equation}

Let $S_{1nk}(t_1)$ be a solution of (\ref{es1}) finite at $t_1=0$, and $S_{2nk}(t_2)$ a solution of (\ref{es2})
vanishing for $t_2 \rightarrow \infty$. Then, up to normalization constants, we have \vspace{-0.1cm}
\begin{eqnarray}\label{ss1}
S_{1nk}(t_1)
&=&
t_1^{|m|/2}(1-t_1)^{-ik/2}  \nonumber\\[2mm]
&\times& {}_2F_1\left(\frac{1+n+\sigma+|m|-ik}{2}, \frac{1-n-\sigma+|m|-ik}{2}; |m|+1; t_1\right),
\\[3mm] \label{ss2}
S_{2nk}(t_2)
&=&
t_2^{(ik-n+\sigma-1)/2}(t_2-1)^{-ik/2} \quad \nonumber \\[2mm]
&\times& {}_2F_1\left(\frac{1+n-\sigma+|m|-ik}{2}, \frac{1+n-\sigma-|m|-ik}{2}; n-\sigma+1; \frac{1}{t_2} \right),
\end{eqnarray}
where $k=\sqrt{A}$ and $\sigma=\alpha R/n$. Both $S_{1nk}$ and $S_{2nk}$ are real functions.
By using equation (\ref{gi}) we can evaluate the following integrals
\begin{eqnarray}
\label{if1}
\int_0^1S_{1nk}^2(t)dt
=\frac{2\pi(|m|!)^2\,\mathrm {Im}\left[\psi\left(\frac{|m|+1-n-\sigma+ik}{2}\right)
- \psi\left(\frac{|m|+1+n+\sigma+ik}{2}\right)\right]} {(n+\sigma)\sinh\pi k \left|\Gamma\left(\frac{|m|+1+n+\sigma+ik}{2}\right)\right|^2
\left|\Gamma\left(\frac{|m|+1-n-\sigma+ik}{2}\right)\right|^2},
\end{eqnarray}
\begin{eqnarray}
\label{if2}
\int_1^{\infty}S_{2nk}^2(t)dt
&=& \frac{2\pi(n-\sigma)[\Gamma(n-\sigma)]^2}{\sinh\pi k}
\nonumber
\\[3mm]
&\times& \frac{\mathrm {Im}\left[\psi\left(\frac{|m|+1+n-\sigma+ik}{2}\right)
+  \psi\left(\frac{-|m|+1+n-\sigma+ik}{2}\right)\right]} {\left|\Gamma\left(\frac{|m|+1+n-\sigma+ik}{2}\right)\right|^2
\left|\Gamma\left(\frac{-|m|+1+n-\sigma+ik}{2}\right)\right|^2},
\end{eqnarray}
\begin{eqnarray}
\int_0^1S_{1nk}(t)S_{1nk^{\prime}}(t)\frac{dt}{1-t}= \frac{4\pi^2(|m|!)^2\delta(k-k^{\prime})}
{k\sinh\pi k \left|\Gamma\left(\frac{|m|+1+n+\sigma+ik}{2}\right)\right|^2
\left|\Gamma\left(\frac{|m|+1-n-\sigma+ik}{2}\right)\right|^2}, \label{ii1}
\end{eqnarray}
\begin{eqnarray}
\int_1^{\infty}S_{2nk}(t)S_{2nk^{\prime}}(t)\frac{dt}{t-1}= \frac{4\pi^2[\Gamma(n-\sigma+1)]^2\delta(k-k^{\prime})} {k\sinh\pi k \left|\Gamma\left(\frac{|m|+1+n-\sigma+ik}{2}\right)\right|^2
\left|\Gamma\left(\frac{-|m|+1+n-\sigma+ik}{2}\right)\right|^2}. \label{ii2} \vspace{-0.1cm}
\end{eqnarray}
Here $\psi(z)=d\ln\Gamma(z)/dz$. Now, denoting
\begin{eqnarray}
\label{function1}
\Psi_{nkm}(t_1,t_2,\varphi)=N_{km}^{n}S_{1nk}(t_1)S_{2nk}(t_2)e^{im\phi}
\end{eqnarray}
and using integrals (\ref{if1}) -- (\ref{ii2}), we obtain the normalization relation
\begin{equation}
\int\!\!\!\!\!\!\int\limits_{x_0+x_3>0}\!\!\!\!\!\!\int\Psi^{\ast}_{nkm}(t_1,t_2,\phi)
\Psi_{nk^{\prime}m}(t_1,t_2,\phi)dV     = \pi\delta(k-k^{\prime}),
\end{equation}
where $dV$ is the volume element in coordinates $t_1,\,t_2,\,\phi$,
$$
dV=\frac{R^3}{4}\left(\frac{1}{1-t_1}+ \frac{1}{t_2-1}\right)dt_1dt_2d\phi,
$$
and the normalization coefficient is
$$
\begin{array}{c}
               {\displaystyle N_{km}^{n}=\frac{\sqrt{k}\sinh\pi k\sqrt{n^2-\sigma^2} }
               {2\sqrt{\pi}R^{3/2}\Gamma(n-\sigma+1)|m|!\sqrt{I(n,\sigma)}} } \\[3mm]
\times  \left|\Gamma\left(\frac{|m|+1+n+\sigma+ik}{2}\right)\right|
\left|\Gamma\left(\frac{|m|+1-n-\sigma+ik}{2}\right)\right|
\left|\Gamma\left(\frac{|m|+1+n-\sigma+ik}{2}\right)\right|
\left|\Gamma\left(\frac{-|m|+1+n-\sigma+ik}{2}\right)\right|.
             \end{array}
$$
Here
$$
\begin{array}{r}
   I(n,\sigma)=
\mathrm {Im}\left\{(n-\sigma) \left[\psi\left(\frac{|m|+1-n-\sigma+ik}{2}\right)- \psi
\left(\frac{|m|+1+n+\sigma+ik}{2}\right)\right] \right. \\[3mm] \left.
  +\,(n+\sigma) \left[\psi\left(\frac{|m|+1+n-\sigma+ik}{2}\right)+ \psi\left(\frac{-|m|+1+n-\sigma+ik}{2}\right)\right]\right\}. \\
\end{array}
$$
Normalization simplifies greatly for $\alpha=0$. In this case,
$$
N_{km}^{n}=\frac{\sqrt{nk\sinh\pi k}\left|\Gamma\left(\frac{|m|+n-ik+1}{2}\right)\right|^2}{\pi R^{3/2}n!}.
$$

For $E=E_n$ the separation constant $A$ may also take discrete values. This is only possible if $n-\sigma<|m|<n+\sigma$.
Then we may write $A=-[\sigma-(n_1+n_2+1)]^2$,
where $n_1=0,\,1,\,2,\,...,\linebreak\frac12[n+\sigma-|m|-1)],\,n_2=0,\,1,\,2,\,...,\frac12[|m|-n+\sigma-1)] $,
and $n=|m|+n_1-n_2$. In this case the wave functions $\Psi_{n_1n_2m} (t_1, t_2, \phi)$ normalized by the condition
\begin{equation}
\int\!\!\!\!\!\!\int\limits_{x_0+x_3>0}\!\!\!\!\!\!\int\Psi^{\ast}_{n_1n_2m} (t_1, t_2, \phi) \Psi_{n_1n_2m} (t_1, t_2, \phi)dV     = \frac12,
\end{equation}
take the form
\begin{equation}\label{psn1n2}
\Psi_{n_1n_2m} (t_1, t_2, \phi)=N_{n_1n_2}^{n}\,
S_{1n_1 n_2}^n (t_1)\,S_{2n_1 n_2}^n (t_2)\,\frac{e^{i m\phi}}{\sqrt{2\pi}},
\end{equation}
where
\begin{eqnarray}
S_{1n_1 n_2}^n (t_1)\!\! &=&t_1^{|m|/2}(1-t_1)^{-n_1+(n+\sigma-|m|-1)/2}\,{}_2F_1(-n_1,\,n+\sigma-n_1;\, |m|+1; \,t_1), \nonumber \\[3mm]
S_{2n_1 n_2}^n (t_2)\!\!&=&t_2^{n_2-|m|/2}(t_2-1)^{-n_2+(|m|-n+\sigma-1)/2}\,{}_2F_1(-n_2,\,|m|-n_2; \, n-\sigma+1; \, 1/t_2), \nonumber
\end{eqnarray}
and
\begin{eqnarray}
N_{n_1n_2}^{n}&=&\frac{\sqrt{(n^2-\sigma^2)(\sigma-n_1-n_2-1)}}{|m|!\Gamma(n-\sigma+1)}
\sqrt{\frac{(|m|+n_1)!(|m|-n_2-1)!}{n R^3 n_1!n_2!}} \nonumber \vspace{-0.1cm}
\\[3mm]
&\times&
\sqrt{\frac{\Gamma(|m|+\sigma-n_2)\Gamma(|m|+n_1-\sigma+1)}
{\Gamma(\sigma-n_1)\Gamma(\sigma-n_2)}} \,.   \nonumber
\end{eqnarray}

For the fixed discrete values of energy $E=E_n$, the wave functions (\ref{function1}) or (\ref{psn1n2})
can be expanded in terms of eigenfunctions $\Psi_{n \ell m} (\tau, \theta, \phi)$ (\ref{psnl}).
Here we consider only the interbasis expansion for the function (\ref{psn1n2}). The case of expansion
of the function (\ref{function1}) is calculated in a similar manner.

Let us write down the desired expansion in form
\begin{equation}
\label{expan}
\Psi_{n_1 n_2 m}(t_1,\,t_2,\phi) =
\sum_{l = \max \{n, |m| \}}^\infty  \, W_{n_1 n_2}^{n \ell m} \,
\Psi_{n \ell m} (\tau, \theta, \phi).
\end{equation}
To find expansion coefficients $W_{n_1 n_2}^{n \ell m}$, we first express coordinates $(t_1,\, t_2)$
on the left side of (\ref{expan}) through spherical coordinates (\ref{imls}):
$$
t_1 = (1-\cos\theta)/(1 + \tanh\tau), \qquad
t_2 = (1+\cos\theta)/(1-\tanh\tau).
$$
Here should be noted that wave function (\ref{function1}) is valid only on the half of space $x_0 + x_3 >0$,
or in form of pseudo-spherical coordinates $\tanh\tau + \cos\theta >0$.

Using now the definition of spherical functions  \cite{Var}
\begin{equation}
Y_{\ell m}(\theta,\varphi)=(-1)^{\frac{m+|m|}{2}}
\left[\frac{2\ell+1}{2}\frac{(\ell-\m)!}{(\ell+\m)!}\right]^{\frac{1}{2}}P_\ell^{\m}(\cos\theta)\,\frac{e^{i m\phi}}{\sqrt{2\pi}},
\end{equation}
and the orthogonality condition of Legendre polynomials $P_l^{\m}(\cos\theta)$
\begin{equation}
\int_{-1}^1P_\ell^{\m}(\cos\theta)P_{\ell'}^{\m}(\cos\theta) \sin\theta d\theta = \frac{2}{2\ell+1}\frac{(\ell+\m)!}{(\ell-\m)!} \,\, 
\delta_{\ell, \ell'},
\end{equation}
it is easy to get
\begin{eqnarray}
\label{1-expan}
W_{n_1 n_2 m}^{n \ell m} \,\, S_{n\ell}(\tau)
=  (-1)^{\frac{m+|m|}{2}}
\left[\frac{2\ell+1}{2}\frac{(\ell-\m)!}{(\ell+\m)!}\right]^{\frac{1}{2}} \, N_{n_1n_2}^{n}
\nonumber\\[3mm]
\times
\int_{-\tanh\tau}^1 \, P_\ell^{\m}(x) \, \, S_{1n_1 n_2}^n \left(\frac{1-x}{1 + \tanh\tau}\right)
S_{2n_1 n_2}^n \left(\frac{1+x}{1-\tanh\tau}\right)  dx,
\end{eqnarray}
where we denote $x= \cos\theta$. To simplicity the calculation of interbasis coefficients $W_{n_1 n_2 m}^{n \ell m}$
taking the limit $\tau \to \infty$ in the both sides of relation (\ref{1-expan}).
By using the asymptotic form of wave functions
\begin{equation}
S_{n\ell}(\tau) = \frac{N_{n\ell}\, f(\tau)}{2R^{3/2} \cosh\tau}
\sim \frac{(-1)^{\ell-n}  e^{(\sigma-n-1)\tau}}{\Gamma\left(n+1 - \sigma\right)}
\sqrt{\frac{n^2-\sigma^2}{n} \frac{(\ell+n)!}{(\ell-n)!} \frac{\Gamma(\ell+1-\sigma)}{\Gamma(\ell+1+\sigma)}},
\nonumber
\end{equation}
\begin{eqnarray}
S_{1n_1 n_2}^n \left(\frac{1-x}{1 + \tanh\tau}\right)  \, S_{2n_1 n_2}^n \left(\frac{1+x}{1 - \tanh\tau}\right)
\sim
\left(\frac{1-x}{2}\right)^{\frac{|m|}{2}}\left(\frac{1+x}{2}\right)^{\sigma - n_1 - 1 - \frac{|m|}{2}}
\nonumber\\[3mm]
\times \, e^{(\sigma-n-1)\tau} \,{}_2F_1\left(-n_1,\,n+\sigma-n_1;\, |m|+1; \,\frac{1-x}{2}\right),
\nonumber
\end{eqnarray}
it is easy to see that the dependence on the variable $\tau$ is eliminated on both sides of formula
(\ref{1-expan}). As a result we arrive at the integral representation for the interbasis coefficients
$W_{n_1 n_2 m}^{n \ell m}$:
\begin{eqnarray}
W_{n_1 n_2 m}^{n \ell m} =  (-1)^{\ell-n + \frac{m+|m|}{2}}  \,\, C_{n_1 n_2 m}^{n \ell} \,\, B_{n_1 n_2 m}^{n \ell},
\end{eqnarray}
where
\begin{eqnarray}
C_{n_1 n_2 m}^{n \ell}
&=&
\frac{\sqrt{(\sigma-n_1-n_2-1)}}{(|m|)!}
\sqrt{\frac{2\ell+1}{2}\frac{(\ell-\m)!}{(\ell+\m)!} \frac{(\ell-n)! \Gamma(\ell+1+\sigma)}
{(\ell+n)! \Gamma(\ell+1-\sigma)}}
\nonumber\\[3mm]
&\times&
\sqrt{\frac{(|m|+n_1)!(m-n_2-1)!}{n_1!n_2!}},
\end{eqnarray}
and
\begin{eqnarray}
B_{n_1 n_2 m}^{n \ell} =
\int_{-1}^1 \,
\left(\frac{1-x}{2}\right)^{|m|/2}\left(\frac{1+x}{2}\right)^{-n_1+ \sigma - 1 - |m|/2}
\nonumber\\[3mm]
{}_2F_1\left(-n_1,\,n+\sigma-n_1;\, |m|+1; \,\frac{1-x}{2}\right) \,\,
P_\ell^{\m}(x) \, \,  dx.
\end{eqnarray}
Taking into account the formula
\begin{eqnarray}
{}_2F_1\left(-n_1,\,n+\sigma-n_1;\, |m|+1; \,\frac{1-x}{2}\right)
&=&  (-1)^{n_1}
\frac{n_1!|m|!}{(n_1+\m)!} \frac{\Gamma(\sigma - n_2)}{\Gamma(n+\sigma - |m|)}
\nonumber\\[3mm]
&\times&
\sum_{k=0}^{n_1} \frac{(n_1)_k (n+\sigma-n_1)_k}{(\sigma - n_1 - n_2)_k \,k!}\l(\frac{1+x}{2}\r)^k
\nonumber
\end{eqnarray}
and the explicit form of Legendre polynomials $P_l^{\m}(x)$:
\begin{equation}
P_\ell^{\m}(x) = \frac{(-1)^{|m|}}{\ell! 2^{\ell}} \frac{(\ell+\m)!}{(\ell-\m)!}\,  (1-x^2)^{- \frac{|m|}{2}}\,
\frac{d^{\ell-|m|}}{dx^{\ell-|m|}} (x^2-1)^{\ell}
\end{equation}
we find after some algebraic calculations that
\begin{eqnarray}
B_{n_1 n_2 m}^{n \ell}
&=&
(-1)^{n_1} \frac{2 n_1! |m|!}{(n_1+|m|)!} \frac{(\ell+\m)!}{(\ell-\m)!}\,
\frac{\Gamma(n+\sigma-n_1-|m|)\Gamma(\sigma-n_1)\Gamma(\sigma-n_1-|m|)}
{\Gamma(n+\sigma-|m|)\Gamma(\sigma-n_1+\ell+1)\Gamma(\sigma-n_1-\ell)}
\nonumber\\[3mm]
&\times&
{}_4F_3\left[\begin{array}{c}
 - n_1, n+\sigma - n_1, \sigma-n_1, \sigma-n_1 - |m|; 1 \\
  n+\sigma - |m| - 2 n_1, \sigma - n_1 + \ell +1, \sigma - n_1 - \ell
\end{array}\right].
\end{eqnarray}
where ${}_4F_3$ is the generalized hypergeometric function \cite{Var, Bai}.
Using now the relation \cite{Bai}
$$
{_4F_3}\left[\begin{array}{c}
  -n,b,c,d,1 \\
  e,f,g
\end{array}\right]=\frac{(f-b)_n(g-b)_n}{(f)_n(g)_n}
{_4F_3}\left[\begin{array}{c}
  -n,b,e-c,e-d,1 \\
  e,b-f-n+1,b-g-n+1
\end{array}\right], \vspace{-0.5mm}
$$
we finally have
\begin{eqnarray}
W_{n_1 n_2 m}^{n \ell m}
&=&
\frac{(-1)^{\ell-n + \frac{m-|m|}{2}}}{(\ell)!(n+|m|)!}\,
\frac{\sqrt{2(\sigma-n_1-n_2-1)(2\ell+1)}}{\sqrt{\Gamma(\ell+1+\sigma)\Gamma(\ell+1-\sigma)}}
\,\, \frac{\Gamma(\sigma-n_1-n_2)(\ell-n_1-n)!}
{\Gamma(\sigma+n_1-n_2)\Gamma(n+\sigma-\ell-n_1)}
\nonumber\\[3mm]
&\times&
\sqrt{\frac{(\ell+\m)!}{(\ell-\m)!}
\frac{n_1! (\ell+n)!(|m|+n_1)!(|m|-n_1-1)!}{n_2! (\ell-n)!}}
\,
\Gamma(\sigma-n_1)\Gamma(\sigma-n_2)
\nonumber\\[3mm]
&\times&
{}_4F_3\left[\begin{array}{c}
 - \ell + n, -\ell+|m|, n - n_1,  n+\sigma-n_1; 1 \\
1+|m| + n, n-\ell -n_1, n + \sigma - n_1 - \ell
\end{array}\right].
\nonumber
\end{eqnarray}
Let us also note that the expansion coefficients $W_{n_1 n_2 m}^{n \ell m}$ can be expressed through the Wilson orthogonal polynomials.

\section*{Acknowledgments}

The work of Yu.A.K., V.S.O., and G.S.P. was supported in part by the Armenian-Belarusian grants Nos. 13RB-035 and Ph14ARM-029 from SCS and FFR.

\end{document}